\documentclass[aps,twocolumn,prl]{revtex4}
\usepackage{amsfonts}
\usepackage{graphicx}
\usepackage{epsfig}
\def\bea{\begin{eqnarray}}
\def\eea{\end{eqnarray}}
\def\be{\begin{equation}}
\def\ee{\end{equation}}

\def\cT{{\cal T}}
\begin{document}
\title{Friedel oscillations and the Kondo screening cloud}
\author{Ian Affleck$^{1}$, L\'aszl\'o Borda$^{2}$ and Hubert
  Saleur$^{3,4}$}
 \affiliation{$^{1}$ Department of Physics and Astronomy, University of British 
Columbia, Vancouver, B.C., Canada, V6T 1Z1\\
$^{2}$ Physikalisches Institut, Universit\"at Bonn, Nussallee 12, 
D-53115 Bonn, Germany\\ 
$^{3}$ Service de Physique Th\'eorique, CEA Saclay,
Gif Sur Yvette, 91191, France\\
$^{4}$ Department of Physics and Astronomy,
University of Southern California, 
Los Angeles, CA 90089, USA}
\date{\today}
\begin{abstract}
We show that the long distance charge density oscillations in a metal induced by a weakly coupled spin-1/2 magnetic impurity 
exhibiting the Kondo effect are given, at zero temperature, by a universal function  $F(r/\xi_K)$ 
where $r$ is the distance from the impurity and $\xi_K$, 
the Kondo screening cloud size $\equiv \hbar v_F/(k_BT_K)$, where $v_F$ 
is the Fermi velocity and $T_K$ is the Kondo temperature.  $F$  
is given by a Fourier-like transform of the T-matrix.  Analytic 
expressions for $F(r/\xi_K)$ are derived in both limits $r\ll \xi_K$ and 
$r\gg \xi_K$ and $F$ is calculated for all $r/\xi_K$ using 
numerical  methods. 
\end{abstract}
\maketitle

The interaction of a single magnetic impurity with the conduction electrons in a metal is 
often described by the Kondo model:
\be H=\sum_{\vec k \alpha}\epsilon (\vec k)\psi^\dagger_{\vec k \alpha}\psi_{\vec k \alpha}
+J\sum_{\vec k,\vec k' \alpha \beta}\psi^\dagger_{\vec k \alpha}{\vec \sigma_{\alpha \beta}\over 2}
\psi_{\vec k'\beta}\cdot \vec S.\label{H}\ee
In general we also include a potential scattering term in the Hamiltonian:
$H\to H+V\sum_{\vec k,\vec k'\alpha}\psi^\dagger_{\vec k\alpha}\psi_{\vec k'\alpha}.$
This model exhibits 
 a remarkable crossover from weak to strong coupling behavior as the energy scale is 
lowered through the Kondo temperature, $k_BT_K\approx {\cal D}\exp (-1/\lambda_0 )$ 
where ${\cal D}$ is an ultra-violet cut-off scale (such as a band width) and $\lambda_0$ is 
the dimensionless bare coupling constant ($=J\nu$ where $\nu$ is the density of states, per spin).
 For a review, see, for example, chapter 4 of \cite{Hewson}.  
The renormalized Kondo coupling, $\lambda (E)$, becomes of O(1) at $E\sim k_BT_K$. 
 While physics at energy scales $E\gg k_BT_K$ is given by weak coupling 
perturbation theory, at $E\ll k_BT_K$, the physics is governed by the strong 
coupling fixed point corresponding to a screened impurity and a $\pi /2$ phase shift 
for the low energy quasi-particles.

 The {\it length} dependence of Kondo physics is 
much less well understood. It is generally expected that physical quantities 
exhibit a crossover at a length scale $\xi_K\equiv \hbar v_F/(k_BT_K)$ (where $v_F$ 
is the Fermi velocity), which is typically in the range of .1 to 1 micron. (We henceforth 
set $\hbar$ and $k_B$ to 1.)
For a review see section 9.6 of \cite{Hewson}. See \cite{Gruner1,GZ,BA} for 
original work on the subject.
 However, 
such a crossover at this long length scale  has never been observed experimentally and 
has sometimes been questioned theoretically \cite{Coleman}.
One way of observing this length scale is through the density oscillations 
around a magnetic impurity \cite{Gruner2,Gruner1,GZ,earlier}. It was pointed 
out in \cite{Gruner1} that these should only approach the standard Friedel 
form at distances $r\gg \xi_K$, with a form at shorter distances 
controlled by the $\cT$-matrix. However, experimental data so far doesn't 
seem to support this expectation \cite{GZ}, yielding much shorter 
characteristic lengths. (See also \cite{earlier,Bergman}.) One purpose here is to present a more complete 
theoretical treatment of these density oscillations, since scanning tunnelling 
microscopy of magnetic ions on metallic surfaces provides a new experimental 
technique by which they might now be measured.
Alternative approaches to observing this fundamental length scale involve 
experiments on mesoscopic structures with dimensions of $O(\xi_K)$ \cite{AS}. 

We focus on the case of an $S=1/2$ impurity, and a spherically symmetric dispersion relation 
(normally $\epsilon (\vec k)=k^2/2m-\epsilon_F$).  We consider this 
model in dimension $D=1$, $2$ or $3$.  

There are two reasons why one might be skeptical that the length scale $\xi_K$ would show 
up in the charge density.  One is the idea of ``spin-charge'' separation in  $D=1$ . 
The Hamiltonian of Eq. (\ref{H}) in any dimension, can be mapped into 
a 1D model by expanding in spherical harmonics and using the fact that only 
the s-wave harmonic interacts with the impurity in the case of a $\delta$-function 
interaction. The low energy degrees of freedom of non-interacting 1D electrons 
can be separated into decoupled spin and charge excitations, using bosonization. 
It is possible to write the Kondo interaction in terms of the spin degrees of freedom 
only and hence, one might expect the charge density to be unaffected by the 
Kondo interaction. The fallacy in this argument is that the charge density 
at location $r$ in the 1D model contains a term 
$\psi^\dagger_{L\alpha}(r)\psi_{R\alpha}(r)\exp (-2ik_Fr)+h.c.$ where $R$ and $L$ label right and 
left movers. Standard bosonization methods imply that this term involves both 
spin and charge bosons: $\sin (\sqrt{2\pi}\phi_c+2k_Fr)\cos(\sqrt{2\pi}\phi_s(r))$. 
This is unlike the term $\psi_{L\alpha}^\dagger (r)\psi_{L\alpha}(r)$ which only involves the charge boson. 

Another reason why one might expect no interesting Friedel oscillations follows 
from consideration of the particle-hole (p-h) symmetric case. This symmetry is 
exact, for example, in a nearest neighbor tight-binding model at 1/2-filling 
with the Kondo coupling occuring at the origin only. Then it can easily 
be proven that $<\psi^\dagger_{j\alpha}\psi_{j\alpha}>=1$ for all sites $j$. 
However, a realistic model always breaks particle-hole symmetry.  This 
can be achieved by taking a non p-h symmetric dispersion relation - for 
instance moving the density away from 1/2-filling in the tight-binding 
model. Alternatively, potential scattering can be included in the model.
Then, p-h symmetry is broken even if the dispersion relation doesn't break it. 

We find  for the density oscillations at zero temperature and $r\gg 1/k_F$:
\bea \rho (r)-\rho_0&\to& {C_D\over r^D}[\cos(2k_Fr-\pi D/2+2\delta_P)F(r/\xi_K)\nonumber \\
&-&
\cos(2k_Fr-\pi D/2)].
\label{result}\eea
Here $F(r/\xi_K)$ is a universal scaling function {\em which is the same for all D}, 
$\delta_P$ is the phase shift at the 
Fermi surface produced by the potential scattering, $C_3=1/(4\pi^2)$, $C_2=1/(2\pi^2)$ and 
$C_1=1/(2\pi )$.
In general, there are non-zero oscillations but they vanish exactly in the p-h symmetry 
case for D=1 where $\delta_P=0$, $k_F=\pi /2$ and $r$ is restricted to integer values, 
corresponding to a tight-binding model at 1/2-filling. In the limit of zero 
Kondo coupling, $F=1$ and we recover the standard formula for Friedel 
oscillations produced by a potential scatterer (in the s-wave channel only). 
For a small bare Kondo coupling, $\lambda_0 \ll 1$, $F(r/\xi_K)$ is close to 1 
at $r\ll \xi_K$ so that the oscillations are just determined by the potential scattering, 
$\propto \cos (2k_Fr-\pi D/2+2\delta_P)-\cos (2k_Fr-\pi D/2)$, vanishing if $\delta_P$ is also zero. 
However, at $r\gg \xi_K$, we find that $F(r/\xi_K)\to -1$ which is equivalent 
to $\delta_P\to \delta_P+\pi /2$.  We again recover the potential scattering 
result but now the phase shift picks up an additional contribution of $\pi /2$ 
from the Kondo scattering. 

To derive these results, following \cite{Gruner1},
 it is convenient to relate the scaling function, $F(r/\xi_K)$ to the ${\cal T}$-matrix, 
$\cT (\omega )$ which has already been well-studied by a number of methods and 
is a universal scaling function of $\omega /T_K$. This can be done using 
the standard formula for the (retarded) electron Green's function:
\be G(\vec r,\vec r',\omega)=G_0(\vec r-\vec r',\omega )+G_0(\vec r,\omega )\cT (\omega )
G_0(-\vec r',\omega ),\label{T}\ee
where $G_0$ is the Green's function for the non-interacting case (with 
$J=V=0$). 
This result is a direct consequence of the assumed $\delta$-function form of 
the Kondo (and potential scattering) interaction. The density 
is obtained from the retarded Green's function by:
\be \rho (r)=-{2\over \pi}\int_{-\infty}^0d\omega~\hbox{Im}~G(\vec r,\vec r,\omega ).\label{Gden}\ee
(The factor of 2 arises from summing over spin.)
The exact non-interacting Green's function is:
\bea  
G_0&=&{-ik_F\over v_F\tilde k}\left[{-i\tilde k\over 2\pi r}\right]^{(D-1)/2}
\exp [i\tilde k r],\ \ (D=1, 3)
\nonumber \\ 
&=&-[k_F/(\pi v_F)]K_0\left[ -i\tilde k r\right], \ \ (D=2)\eea
 where 
$\tilde k\equiv \sqrt{k_F^2+2k_F\omega /v_F}$ and $K_0(z)$ is the modified Bessel function. This gives the asymptotic behavior at 
$r\gg 1/k_F$, $\omega \ll {\cal D}$:
\be G_0^2(r,\omega )\to -{1\over v_F^2}\left[{-ik_F\over 2\pi r}\right]^{D-1}\exp [2ik_Fr+2i\omega r/v_F].\label{G0}\ee
(This asymptotic behavior holds for general dispersion relations.) 
The $\cT$ matrix in D-dimensions can be written at $\omega \ll {\cal D}$: $\cT(\omega )=t(\omega /T_K)/(2\pi \nu_D)$
where $t$ is a universal dimensionless function of $\omega /T_K$ 
and $\nu_D$, the density of states per spin at the Fermi energy, has the value 
$\nu_D=k_F^{D-1}/(c_Dv_F)$, with $c_3=2\pi^2$, $c_2=2\pi $ and $c_1=\pi$.
Note that $G_0\cT G_0$ is proportional to the difference between the s-wave Green's function with 
and without the Kondo and potential scattering interactions, since the other spherical 
harmonics are unaffected by the interactions and cancel in $G-G_0$. The effect of the 
s-wave potential scattering at long distances is just to multiply the s-wave Green's 
function by the phase $e^{2i\delta_P}$, thus giving:
\be t(\omega /T_K)=e^{2i\delta_P}[t_K(\omega /T_K)+i]-i,\label{tK}\ee
where $t_K(\omega /T_K)$ is the part of the $t$-matrix coming from the Kondo scattering. 
Combining Eqs. (\ref{T}-\ref{tK}) gives:
\bea \rho (r)&-&\rho_0\to \{ c_D/[\pi^2v_F(2\pi r)^{D-1}]\} \hbox{Im}~\Biggl\{(-i)^{D-1}e^{2ik_Fr}
                \nonumber \\
&\times& \int_{-\infty}^0d\omega e^{2i\omega r/v_F}\left[(t_K(\omega /T_K)+i)e^{2i\delta_P}-i\right]\Biggr\}.\eea
Essentially this formula (for $D=3$ only) was derived in \cite{Gruner1} except that our treatment 
of p-h symmetry breaking is quite different.  Furthermore we apply much more complete 
knowledge of the $\cT$-matrix.
We expect this formula to be valid whenever $\xi_K$, $r\gg 1/k_F$, regardless of the ratio $r/\xi_K$. 
The function $t_K(\omega /T_K)$ is determined from the p-h symmetric Kondo interaction and so 
obeys: $t_K^*(\omega /T_K)=-t_K(-\omega /T_K)$.  Furthermore $t(\omega /T_K)$ is analytic  
 in the upper half complex $\omega$ plane since it is obtained from 
the retarded Green's function. It then follows that $\int_{-\infty}^0 d\omega \exp (2i\omega r/v_F)t_K(\omega /T_K)$ 
is purely real.  A rescaling of the integration variable implies that we may write:
\be \int_{-\infty}^0d\omega e^{2i\omega r/v_F}t_K(\omega /T_K)\equiv [v_F/(2r)][F(rT_K/v_F)-1],\label{F}\ee
where the universal scaling function $F$ is purely real. Thus:
\bea &\rho& (r)-\rho_0\to \{ c_D/[ 2\pi^2(2\pi )^{D-1}r^D]\} \nonumber \\
&&\times \hbox{Im}~\left\{(-i)^{D-1}e^{2ik_Fr}
\left[F(r/\xi_K)e^{2i\delta_P}-1\right]\right\},\label{rho}\eea
giving the result announced in Eq. (\ref{result}). While this derivation assumed 
that the Kondo interaction is a spatial $\delta$-function leading to 
the simple result, Eq. (\ref{T}), we expect our asymptotic formula for $\rho (r)$ to 
be much more generally true, at length scales large compared to the range 
of the Kondo interaction.  

A perturbative calculation of the $\cT$-matrix \cite{Kondo} gives:
\be t_K(\omega ) = -(3i\pi^2/8)[\lambda_0^2+\lambda_0^3\ln ({\cal D}/\omega )^2+\ldots ],\ee
where ${\cal D}$ is of order the ultraviolet cut-off.  The quantity in brackets can be recognized as the first 2 terms 
in the expansion of 
the square of the running coupling 
$\lambda^2 (\omega )$.  
For $\omega \gg T_K$, $\lambda (\omega )\to 1/\ln (|\omega |/T_K)$, 
so one expects $t_K\to -3\pi^2i/[8\ln^2(|\omega |/T_K)]$. Substituting the perturbative expansion into 
Eq. (\ref{F}), gives:
 $F(r/\xi_K)=1-(3\pi^2/8)[\lambda_0^2+2\lambda_0^3\ln (r/a)+\ldots ]$, 
where $a$ is a short distance cut-off of order $v_F/{\cal D}$. Again, we recognize the first terms 
in the expansion of $\lambda^2(r)$, implying the short distance behavior:
\be F(r/\xi_K)\to 1-3\pi^2/[8\ln^2(\xi_K/r)],\ \  (r\ll \xi_K).\label{Fsmall}\ee

It is an interesting fact that $F(r/\xi_K)$ is apparently given by renormalization group 
improved perturbation theory for $r\ll \xi_K$.  This is quite unlike the situation 
for a related quantity, the Knight shift \cite{BA}. This is again 
given by a scaling function, $\chi (r)=(1/T_K)f(r/\xi_K)$ at zero temperature.  However, in this case 
the term of O($\lambda_0^3$) has a coefficient which diverges as the temperature $T\to 0$, even at a fixed small $r$.
This means that the Knight shift at short distances ($r\ll \xi_K$) is {\it not}
given by renormalization group improved perturbation theory, unlike the Friedel oscillations. 
Instead, the Knight shift  exhibits a non-perturbative behavior, even at short distances. 
A conjecture was made for this non-trivial short distance behavior in \cite{BA}. 
The fact that $F(r/\xi_K)$ is perturbative at small $r/\xi_K$ seems to 
follow from the fact that  $\cT (\omega /T_K)$ is perturbative at large $\omega /T_K$ 
together with Eq. (\ref{F}), which presumably implies that the short distance 
behavior of $F$ is given by the high-frequency behavior of $\cT (\omega /T_K)$. 
The general question of which quantities are perturbative or non-perturbative 
at short distances in the Kondo model (and other quantum impurity models) 
remains open.

Perturbation theory for the Friedel oscillations breaks down at $r$ of O($\xi_K$) but at 
$r\gg\xi_K$ we may use Nozi\`eres local Fermi liquid theory. This gives the $\cT$-matrix:
$t_K\to -i[2+i\omega /T_B-3\omega^2/4T_B^2+\ldots ]$. 
Here $T_B$ corresponds to a particular definition of the Kondo temperature.  (See, for 
example, chapter 4 of [\onlinecite{Hewson}].) It is related to the Wilson definition, called simply 
$T_K$ in \cite{Hewson} by $T_B=2T_K/(\pi w)$ with the Wilson number $w\approx .4128$.
Substituting in Eq. (\ref{F}) gives:
\bea F(r/\xi_K)&\to& -1+\pi w\xi_K/(4r)-3(\pi w)^2\xi_K^2/(32r^2)+\ldots \nonumber \\
  &&(r\gg \xi_K),\label{Flarge}\eea
where we have defined $\xi_K$ precisely in terms of the Wilson definition of $T_K$: $\xi_K\equiv v_F/T_K$. 
Nozi\`eres' perturbation theory can be turned into a full perturbation
theory \cite{Lesage} by taking into account  more irrelevant operators in the vicinity of
the low energy fixed point, which give  higher order terms in 
Eq.(\ref{Flarge}).

In order to strengthen our analytical results, we have performed extensive
numerical renormalization group (NRG) calculations \cite{Wilson,Bullareview}. 
In Wilson's NRG technique
--after a logarithmic discretization of the conduction electron band--
one maps the original Kondo Hamiltonian to a semi-infinite chain with the
impurity at the end. As a direct consequence of the logarithmic discretization
the hopping amplitude along the chain falls off exponentially.  This separation of energy
scales allows us to diagonalize the chain Hamiltonian iteratively in order to
approximate the ground state and the excitation spectrum of the full chain.
If one is interested in spatial correlations, however, some care is needed.
The cornerstone of the model, the logarithmic discretization causes
not only the exponential fall-off of the hopping amplitude, but also
a very poor spatial resolution away from the impurity. To tackle that problem,
we introduce Wannier states centered both around the impurity and the point
of interest $r$ thus reducing the problem to a two impurity type calculation.
Such an approach has been demonstrated to work recently by evaluating 
the spin-spin correlation function around a Kondo impurity -see \cite{SC_NRG}. 
To get the amplitude of the charge oscillations one needs the
explicit value of $k_F$ which we obtained by calibrating the 
NRG code with a pure potential scattering model. 
\begin{figure}
\includegraphics[width=8truecm,clip]{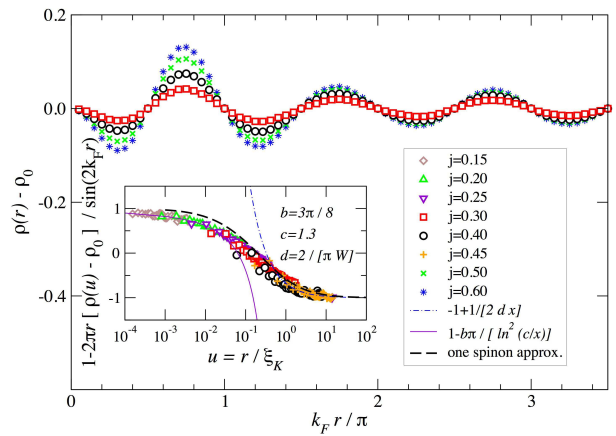}
\caption{
NRG results on charge oscillations around a Kondo impurity
coupled to 1D conduction electrons with particle-hole symmetry. Note that the oscillations vanish at
$k_Fr/\pi\in\mathbb{N}$. As shown in the inset,
the properly rescaled envelope function of the oscillations  
(extracted as $\rho-\rho_0$ at the local maxima)
for different Kondo couplings
nicely collapse into one universal curve except for the points
where $r\sim k_F^{-1}$. 
In the inset we show the analytical results
for the asymptotics as well: Note the good agreement between the analytical
results and the numerics.
}
\label{fig:NRG}
\end{figure}

We show results for different Kondo couplings in Fig. (\ref{fig:NRG}). 
$\rho(r)-\rho_0\sim\sin(2k_Fr)$  in 
agreement with Eq. (\ref{result}) for $\delta_P=0$, the expected 
p-h symmetric result since we use a flat symmetric band with no
potential scattering.
In the inset of Fig.\ref{fig:NRG} we show NRG results for 
$F(r/\xi_K)$ showing 
good agreement with the asymptotic predictions of Eqs. 
(\ref{Fsmall}) and (\ref{Flarge}) and fair agreement with
 the prediction of the ``one spinon approximation''~\cite{Gruner1,LS}, $t_K=-2i/[1-i\omega /T_B]$,
$F(u)=1+4uae^{2ua}\hbox{Ei}~(-ua)$, $a=T_B/T_K=2/(\pi w)\approx 1.542$. (Ei is the 
exponential-integral function.) 
This is a challenging NRG calculation 
since universal behavior is only expected to occur for $\xi_K$, $r\gg k_F^{-1}$ 
(i.e. at distances beyond several periods of the density oscillation and at weak coupling). On the 
other hand, the numerical error increases at large $r$. 
The non-universal, coupling dependent part of the charge density oscillations
is much more extended in space than that of the spin-spin correlator computed 
in \cite{SC_NRG}.  That is the main source of the scattering of data points in the 
inset of Fig. (\ref{fig:NRG}). 
It is interesting to note, from the figure, that $F\approx 0$, corresponding 
to the midpoint of the crossover from weak to strong coupling, occurs at 
$r\approx (0.12\pm0.02)\xi_K$.
Thus an experimental detection of the Kondo screening cloud via the density oscillations would 
``only'' need to measure out to distances of order $\xi_K/10$ to see at least half  of 
the crossover. In STM experiments the most readily accessible measure of the Kondo temperature 
is the half-width of Im $\cT (\omega )$, $T_{1/2}\approx 2T_K$ \cite{Bullareview}. 
Once this number is determined experimentally, 
then the midpoint of the cross-over of the Friedel oscillations is predicted to occur at 
$r\approx v_F/[5T_{1/2}]$.  At finite temperature, Friedel oscillations decay exponentially 
with a thermal correlation length $\xi_T\equiv 2\pi v_F/T$ so it is necessary to 
be at sufficiently low $T$ that $\xi_K>\xi_T$ to measure the Kondo screening cloud. 
Direct electron-electron interactions, ignored in the Kondo model, can also lead
to  decay of the Friedel oscillations with a decay length related to 
the inelastic scattering length.  However, 
Fermi liquid theory 
(typically believed to be valid in D=2 or 3) implies that this length also 
diverges as $T\to 0$.

The Kondo screening cloud {\it does not} show up in the 
energy resolved density of states, $-(2/\pi )$ Im $G(\vec r,\vec r,\omega )$, measured
in STM and given 
by Eq. (\ref{T}). This has a trivial $r$-dependence $1/r^{D-1}$ at $r\gg 1/k_F$. 
At fixed $r$ the Kondo scale only enters through the $\omega$-dependence. Only 
after doing the $\omega$-integral, to get the total electron density does 
the Kondo scale appear in the $r$-dependence. 

Previous attempts \cite{GZ}
 to fit experimental data on density oscillations around Cu and Mn impurities in Al 
to formulas like Eq. (\ref{result}) have yielded characteristic lengths which are much smaller than $\xi_K$ as 
determined from the experimentally measured Kondo temperature.  We think these issues deserve revisiting, using STM. 
Im $\cT$, measured from the energy-resolved density of states (at a fixed 
location near the impurity), has a peak with a width identified as $T_K$. 
This identification is not completely obvious since it is 
typically not feasible to raise the temperature past $T_K$ (due to 
diffusion of the impurity) nor to apply magnetic fields corresponding 
to Zeeman energies of O($T_K$).  
It follows from Eq. (\ref{Gden}) that there should be a change 
in the envelope of the density oscillations at the corresponding 
length scale $v_F/T_K$. An accurate measurement of $\rho (r)$, if 
it agrees with our results,  would both resolve an open 
fundamental question in Kondo physics and firmly 
establish that these systems really do exhibit the Kondo effect. 
  We emphasize that the large 
size of the Kondo cloud  makes 
it very hard to observe. At such large distances that $F(r/\xi_K)$ has changed 
significantly from its short distance asymptote of one, the $1/r^D$ factor in Eq. (\ref{result})
makes the oscillations very small. Clearly the situation is improved in 2 dimensional systems. 

In conclusion, we have shown that the Friedel oscillations around 
a Kondo impurity exhibit a universal behavior characterized by 
the length scale $\xi_K$. We have determined the corresponding universal scaling function 
analytically in both limits $r\ll \xi_K$ and $r\gg \xi_K$ and numerically 
at intermediate $r/\xi_K$. It exhibits renormalization group improved weak coupling 
behavior at short distances, quite unlike the Knight shift, raising intriguing  
general questions about which quantities are perturbative and which are not in this limit,  
for this and other models. 
The envelope of the oscillations, given in Eq. (\ref{result}), 
exhibits a crossover from short to long distances corresponding to an increase of the s-wave 
phase shift by $\pi /2$.  However, at intermediate distances, the result {\it does not} correspond 
to simple potential scattering for {\it any} value of the phase shift. We have 
determined precisely the distance at which the crossover occurs in terms of the measure of 
the Kondo temperature accessible to STM experiments. 

We thank L. Ding, Y. Pennec and A. Zawadowski for helpful discussions. 
This research is supported in part by NSERC and CIfAR (IA),
by the J\'anos Bolyai Foundation, 
the Alexander von Humboldt Foundation
and 
Hungarian Grants OTKA
through projects K73361 and T048782 (LB) and
by  the ESF program
INSTANS (HS).

\end{document}